\documentclass{article}
\usepackage{spconf,amsmath,graphicx}
\usepackage[ruled,vlined]{algorithm2e}
\usepackage{spconf,amsmath,graphicx,multirow,tabularx, booktabs,array}

\title{improving RNN transducer with normalized jointer network}
\name{Mingkun Huang, Jun Zhang, Meng Cai, Yang Zhang, Jiali Yao, Yongbin You, Yi He, Zejun Ma}
\address{Bytedance AI-Lab \\
{\small \tt huangmingkun@bytedance.com}}
\begin{document}
%
\maketitle
\begin{abstract}
Recurrent neural transducer (RNN-T) is a promising end-to-end (E2E) model in automatic speech recognition (ASR). It has shown superior performance compared to traditional hybrid ASR systems. However, training RNN-T from scratch is still challenging. We observe a huge gradient variance during RNN-T training and suspect it hurts the performance. In this work, we analyze the cause of the huge gradient variance in RNN-T training and proposed a new \textit{normalized jointer network} to overcome it. We also propose to enhance the RNN-T network with a modified conformer encoder network and transformer-XL predictor networks to achieve the best performance. Experiments are conducted on the open 170-hour AISHELL-1 and industrial-level 30000-hour mandarin speech dataset. On the AISHELL-1 dataset, our RNN-T system gets state-of-the-art results on AISHELL-1's streaming and non-streaming benchmark with CER 6.15\% and 5.37\% respectively. We further compare our RNN-T system with our well trained commercial hybrid system on 30000-hour-industry audio data and get 9\% relative improvement without pre-training or external language model.



\end{abstract}
\section{Introduction}
\label{sec:intro}

Lots of progress has been made in automatic speech recognition fields in recent years. Many new neural networks such as Deep Neural Network (DNN), Convolution Neural Network (CNN), Recurrent Neural Network (RNN), and Self-attention Neural Network (SAN) \cite{NIPS2017_7181} have greatly boosted the acoustic model performance for hybrid ASR models. However, the hybrid ASR model has a long pipeline that has to be built independently. Recent advances in end-to-end(E2E) models \cite{chan2016listen, prabhavalkar2017comparison, battenberg2017exploring, sak2017recurrent} reveal the E2E model has great potential to replace the traditional hybrid model in the industry area. The E2E ASR system directly transduces input utterance to a sequence of readable tokens. Among the E2E approaches(CTC \cite{graves2006connectionist}, LAS \cite{chan2016listen}, RNN-T \cite{graves2012sequence}), RNN-T has a few advances. First, it has great streaming nature compared to LAS models. Second, it has internal language models compared to CTC models, which learn the acoustic and language features jointly. Last but not least, RNN-T can recognize very long utterance stably \cite{chiu2019comparison}. Recently, Google has successfully deployed RNN-T on their mobile devices for streaming ASR \cite{he2019streaming}, which greatly impacts both the academic and industrial areas.

However, training RNN-T from scratch is still a challenging issue and is actively studied in recent years \cite{irie2019choice,chai2019investigation,sainath2020streaming,mohamed2019transformers,huang2020conv}. Rao et al. revealed that RNN-T system converges better with an encoder network pre-trained with CTC and predictor network pre-trained with language model \cite{rao2017exploring}. Tian et al. improved RNN-T performance by adding alignment constraints to RNN-T's loss function \cite{tian2020synchronous}. Hu et al. showed RNN-T's encoder network with CE pre-train performs better than other pre-train methods \cite{hu2020exploring}. These systems still have a long pipeline due to the need for pre-training. Most recently, advanced networks such as Contextnet\cite{han2020contextnet} and Conformer\cite{gulati2020conformer} have been introduced to RNN-T systems, which achieve SOTA performance on Librispeech benchmark. However, the detail of training is not yet well explained. We further exploited the possibility of leveraging these networks in the RNN-T system without the requirement for pre-training. 

\label{sec:rnnt}
\begin{figure}
\centering
\centerline{\includegraphics[width=0.7\columnwidth]{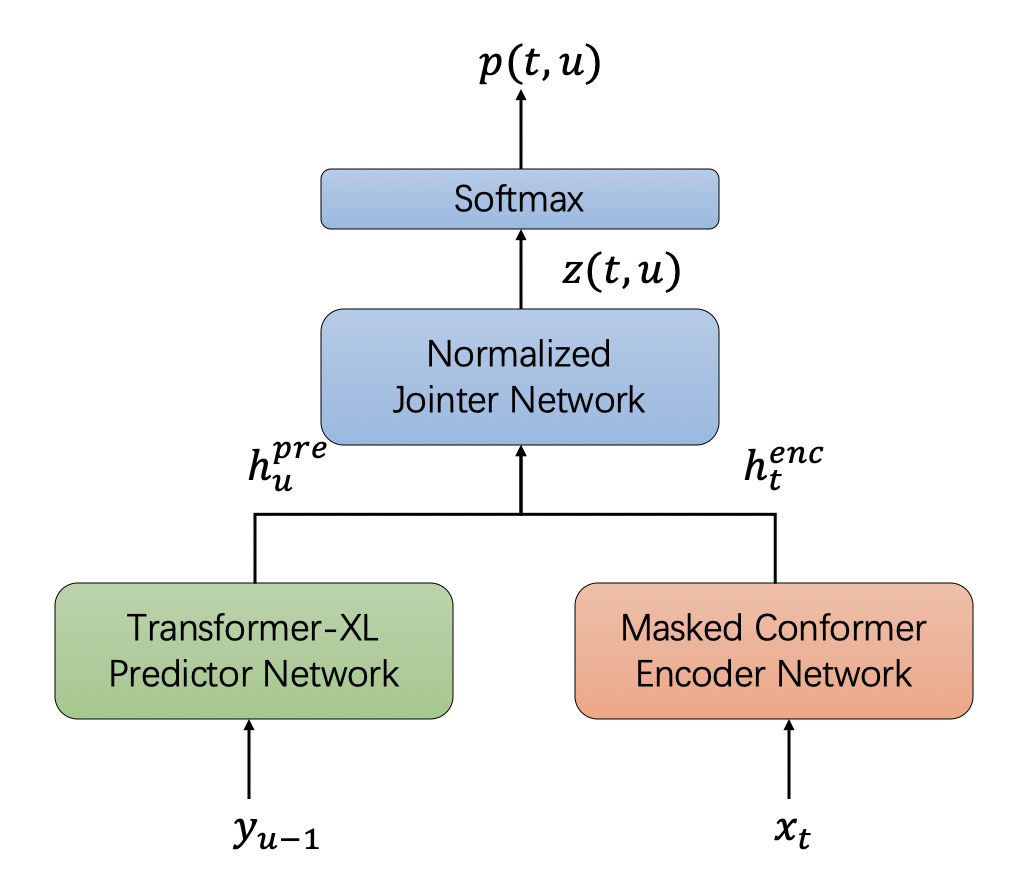}}
\caption{Our RNN-T baseline system consists of a masked conformer encoder network, a transformer-XL predictor, and a proposed normalized jointer network. }
\label{fig1}
\end{figure}

\begin{figure}
\centering
\centerline{\includegraphics[width=0.8\columnwidth]{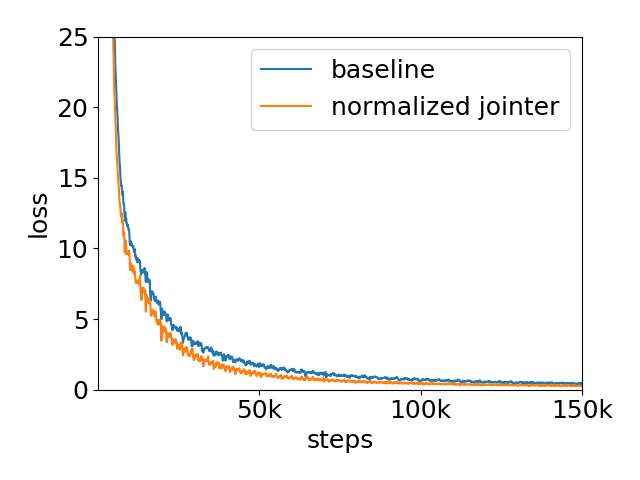}}
\caption{The validation loss curve with our proposed normalized jointer network.}
\label{fig2}
\end{figure}

In this work, we observed a huge gradient variance during RNN-T training and suspect this prevents RNN-T from fast convergence. We analyze the cause of the huge variance in RNN-T training. We find the issue arises during the back-propagation from the jointer to predictor and encoder networks. For the encoder network, the gradient variance is amplified for \(U\) (transcription length) times. While the gradient variance for predictor is amplified for \(T\) (acoustic length) times. This makes the predictor and encoder hard to be optimized. To address the issue, we propose the \textit{normalized jointer network} which applies normalization for gradients at both encoder and predictor by a factor of \(U\) and \(T\) respectively. The experiment shows that the approach improved the performance and accelerated the convergence of the training procedure. Furthermore, we propose to use a masked Conformer as the and encoder network and the Transformer-XL \cite{dai2019transformer} as the predictor network to further improve the performance of the RNN-T system. 

With all the proposed improvements, we achieve SOTA in the Chinese AIShell-1 dataset. In the AISHELL-1 dataset,  we get CER 6.15\% and 5.37\% for streaming and non-streaming recognition respectively. We also compared our improved RNN-T system with an internal well-trained hybrid system based on 30000-hour-industry data. The improved RNN-T system outperforms the hybrid system even without external language models. 

The contribution of this work is threefold. First, we observe the huge gradient variance issue and propose a normalized jointer network that addresses the issue. Second, we further exploit the potential RNN-T system with a proposed masked conformer encoder network and Transformer-XL predictor network. Third, the experiment result outperforms the SOTA of AISHELL-1 with the proposed RNN-T network.

\section{RNN-T System Description}
Figure \ref{fig1} shows the diagram of the RNN-T model. The encoder network is analogous to the encoder networks in LAS \cite{chan2016listen}, or the acoustic model in hybrid system. Usually the input to encoder network is mel-fbank features \(X=[x_1,x_2,...,x_t,...,x_T]\), the encoder network converts the mel-fbank features to the high-level representations \(H^{enc}=[h_1^{enc},h_2^{enc},...,h_t^{enc},...,h_T^{enc}]\). The encoder network can be modeled by LSTM, CNN, DFSMN \cite{zhang2018deep} or transformer networks \cite{NIPS2017_7181}. Lots of works have done to improve the performance of ASR system by modifying the encoder network \cite{han2020contextnet, gulati2020conformer}. The predictor network plays the role of language model for RNN-T system. The non-blank tokens \(Y=[y_1,y_2,...,y_u,...,y_U]\) are encoded by predictor to high-level representations \(h^{pre}=[h_1^{pre},h_2^{pre},...,h_u^{pre},...,h_U^{pre}]\). The predictor networks are normally LSTM, causal CNN or causal transformer networks \cite{NIPS2017_7181}. The jointer network combines the high representation from encoder and predictor network as follows
\begin{equation}
\label{eq1}
    z(t,u)=FC(tanh(h_t^{enc} + h_u^{pre}))
\end{equation}
where FC(.) is the fully connected network. The posterior  \(p(t,u)\) can be obtained by applying softmax function to \(z(t,u)\)
\begin{equation}
\label{eq2}
    p(t,u)=softmax(z(t,u))
\end{equation}

and the RNN-T loss is defined as 
\begin{equation}
\label{eq3}
    L=\sum_{(t,u):t+u=T+U}{\alpha(t,u)\cdot\beta(t,u)}
\end{equation}
where 
\[\alpha(t,u)=\alpha(t-1,u)\cdot p^{blk}(t,u)+\alpha(t,u-1)\cdot p^{y_{u-1}}(t,u-1)\]
\[\beta(t,u)=\beta(t+1,u)\cdot p^{blk}(t,u)+\beta(t,u+1)\cdot p^{y_u}(t,u)\]
Since \(\alpha(t,u)\) and \(\beta(t,u)\) can be effectively computed by forward-backward algorithm \cite{graves2012sequence}, the whole system can be trained in an end to end manner.

\section{Improvement to RNN-T System}
In this section, three improvements to our RNN-T system are detail explained.

\subsection{Masked Conformer Encoder}
\label{ssec:masked_conformer}
Our RNN-T baseline adopts conformer \cite{gulati2020conformer} network as the encoder. The conformer network adds convolution to SAN's  \cite{NIPS2017_7181} self-attention part. We modify the conformer network by adding mask to self-attention part. The self-attention output at index \(i\) can be calculated as Eq.\ref{eq4}. For more details of self-attention mechanism, please refer to \cite{NIPS2017_7181}.
\begin{equation}
\label{eq4}
out_i = \sum_{l} {\left( \cfrac {\exp (\cfrac {Q_i \cdot K_l} {\sqrt {d_k}} ) \cdot M_{il} } { \sum_{j}  {\exp ( \cfrac {Q_i \cdot K_j}  {\sqrt {d_k}} )} \cdot M_{ij}}  \cdot V_l  \right)}
\end{equation}
This mask mechanism introduced in conformer has two advantages. First, adding mask to self-attention in conformer helps the convergence, especially when the training utterance is very long. We observed that the masked conformer always converges faster and are more stable than the original conformer networks. Second, with the mask introduced in self-attention, it is quite easy to change a non-streaming RNN-T system to a streaming fashion by masking out the right context of self attention part.

\subsection{Transformer-XL Predictor}
\label{ssec:transformerxl_predictor}
The RNN-T predictor networks play the role of the neural language model and usually adopts LSTM or native causal transformer network. However, as far as we can see, transformer-XL \cite{dai2019transformer} performs the best in the language model domain. The transformer-XL contains a segment-level recurrence mechanism, which maintains an extra-long context. Meanwhile, transformer-XL proposed a novel positional encoding scheme that adapts the sinusoid formulation in the relative positional embedding. It helps the model generalize to a longer length during evaluation. So we choose transformer-XL as our RNN-T's predictor networks and find it works better than the LSTM or transformer predictor networks. As the training corpus size increases, the RNN-T system with the transformer-XL predictor network will results in much better performance than LSTM or native transformer predictor network.

\subsection{Normalized Jointer}
\label{ssec:normalized_jointer}
The jointer network composes the hidden feature from encoder and predictor network in Eq.\ref{eq1}. In the forward progress computation, the jointer network representation \(z(t,u)\) has the following relationship with \(h_t^{enc}\) and \(h_u^{pre}\)
\begin{equation}
\label{eq5}
    z(t,:) = FC(tanh((h_t^{enc}+h_{:}^{pre})) 
\end{equation}
\begin{equation}
\label{eq6}
    z(:,u) = FC(tanh((h_{:}^{pre}+h_u^{pre}))
\end{equation}
In the backward progress, gradient of \(dh_t^{enc}\) and \(dh_u^{pre}\) will have the following relationship with jointer network's gradient \(dz(t,u)\)
\begin{equation}
\label{eq7}
    dh_t^{enc}=\sum_{u=1}^{u=U}{dz(t,u)}
\end{equation}
\begin{equation}
\label{eq8}
    dh_u^{pre}=\sum_{t=1}^{t=T}{dz(t,u)}
\end{equation}

Let the gradient of \(dz(t,u)\) be a random variable with mean \(\mu\) and variance \(\sigma\). Then the variance of \(dh_t^{enc}\) is \(U\cdot\sigma\), and the variance of \(dh_u^{pre}\) is \(T\cdot\sigma\).  The huge variance of the gradient to RNN-T's encoder and predictor network will cause the unhealthy optimization of parameters, especially when the training batch utterances have different acoustic length (T) and different transcript length (U) in a mini-batch. To overcome this problem, we apply a simple method by dividing the \(dh_t^{enc}\) with \(U\) and \(dh_u^{pre}\) with \(T\). As we can see in Eq.\ref{eq9} and Eq.\ref{eq10}, 

\begin{equation}
\label{eq9}
    \hat{dh_t^{enc}}=\frac{dh_t^{enc}}{U}
\end{equation}
\begin{equation}
\label{eq10}
    \hat{dh_u^{pre}}=\frac{dh_u^{pre}}{T}
\end{equation}

with the above modification to the gradient from RNN-T's jointer network to encoder and predictor network, the gradient norm of RNN-T's training becomes more stable. And the validation loss decreases faster and lower as illustrated in Figure\ref{fig2}.

\section{Experiments}
In this section, we conduct two mandarin speech recognition tasks, named AISHELL-1 task and 30000-hour-industry task. In the AISHELL-1 task, we carry out our experiments on the 170-hour AISHELL-1 dataset \cite{bu2017aishell}. In the 30000-hour-industry task, we collect 30,000 hours of audio data from various acoustic domains and content domains.   The acoustic domain contains utterances collected from fields such as \textit{near field record}, \textit{far field record}, and \textit{telephone record}. The content are from domains such as \textit{news}, \textit{meeting}, \textit{recreation}, \textit{sports}, \textit{tourism}, \textit{game}, \textit{literature}, \textit{education}. We adopt raw waveform as our input and compute the 80-dimensional energy-based log-mel filter-banks (FBK) on-the-fly. The filter-banks are computed on a window of 25ms and 10ms shift. All our encoder networks have VGG blocks with two convolutions and down-sampling the FBK acoustic feature to a frame rate of 40ms. All our experiments are carried with Pytorch and our RNN-T system is trained from scratch without any auxiliary loss or pre-train methods, which greatly simplify the pipeline of this system.

\begin{table}[t]
\small
  \centering
  \label{table1}
  \begin{tabular}{ccccc}
    \toprule
    \multicolumn{1}{c}{\textbf{ID}} &
    \multicolumn{1}{c}{\textbf{Model}} & 
    \multicolumn{1}{c}{\textbf{Latency}} & 
    \multicolumn{1}{c}{\textbf{Param.}}  & 
    \multicolumn{1}{c}{\textbf{Testset}}\\
    \midrule
    -&Path-Aware SA-T \cite{tian2019self} &inf&  - & 9.30  \\
    \midrule
    -&SAN-M \cite{gao2020san} &  inf&43M & 6.46\\
    \midrule
    B0&baseline& inf&61M & 6.35\\
    \midrule
    E1&+Transformer-XL& inf & 46M &  6.18\\
    \midrule      
    E2&+Masked Conformer & inf& 46M &  6.09\\
    \midrule   
    E3& + Normalized Jointer&inf & 46M &  \textbf{5.91}\\
    \midrule
    E4&+ Large &  inf &110M&  \textbf{5.37}\\
    \midrule    
    \midrule
    -&Sync-Transformer \cite{tian2020synchronous} & 400ms & - &  8.91\\
    \midrule
    -&SCAMA \cite{zhang2020streaming} & 600ms & 43M & 7.39\\
    \midrule
    E5& Streaming Base & 400ms & 46M &  \textbf{6.83}\\
    \midrule
    E6& Streaming Large & 400ms & 110M & \textbf{6.15}\\
    \midrule
  \end{tabular}
  \caption{Experiment results on AISHELL-1 task.}
\end{table}

\subsection{AISHELL-1 Task}
We first evaluate the performance of our RNN-T system on the AISHELL-1 benchmark. We build our baseline RNN-T system with 12 layers of default conformer encoder \cite{gulati2020conformer}, and set multi-head number as 4, dimension as 256. The fully connected hidden size is 2048. The predictor network is a 1 layer LSTM with cell size 2048. The jointer network has a dimension of 768, while the output vocabulary is 4234. With this configuration, the baseline RNN-T system has 61M parameters (\textit{B0}).  We use Adam \cite{kingma2014adam} as the optimizer with \(\beta_1=0.9\), \(\beta_2=0.999\). The learning rate is set to rise linearly from 1e-7 to 5e-4 in the first 15,000 steps, then exponentially decay it in the rest of our experiments. Our RNN-T system is trained on the AISHELL-1 task with 128 utterances per mini-batch, the model is stopped after 150,000 iterations. The dropout is set to 0.1 in all the experiments. The SpecAugment \cite{park2019specaugment} is adopt as default. All the results are reported without any external language model.

As shown in Table 1, our baseline RNN-T system (\textit{B0}) gets CER 6.35\% on the AISHELL-1 testset. It outperforms the previous state-of-the-art benchmark achieved by SAN-M \cite{gao2020san}. After replacing the LSTM predictor network in \textit{B0} with the transformer-XL predictor, the CER decreases to 6.18\% with this modification. The parameters of our RNN-T system are reduced from 61M to 46M (\textit{E1}). Then, we improved our RNN-T system by replacing the default conformer encoder network with the proposed masked conformer encoder network, where each layer has a mask with 40 left contexts and 40 right contexts. With masked conformer encoder network, the CER drops to 6.09\% on testset (\textit{E2}). In \textit{E3}, we adopt our proposed normalized jointer network to \textit{E2} and observe a signification CER reduction from 6.09\% to 5.91\%. Figure \ref{fig2} presents our proposed jointer network helps the RNN-T system converging faster and better. In \textit{E4}, we further increase our RNN-T's model complexity by set the multi-head to 8, and dimension to 512. With this bigger RNN-T model (110M parameters), we get CER 5.37\% on the AISHELL-1 testset. It further improves the state-of-the-art benchmark.

For the streaming RNN-T system, we simply set the right context of mask in the first 10 encoder layers to 1, and 0 in the remaining layers. In this setting, the streaming RNN-T System has a 400ms latency. We denote it \textit{Streaming Base}. This system gets CER 6.83\% on the AISHELL-1 testset, which again outperforms the previous state-of-the-art benchmark on AISHELL-1 (\textit{E5}). We further conduct our streaming RNN-T experiment with our 110M parameters system, denoted as \textit{Streaming Large}. It gets CER 6.15\% on the AISHELL-1 testset, which again improves the streaming benchmark on AISHELL-1 (\textit{E6}).

\subsection{30000-hour-industry Task}
In this task, we carry out our experiments on our internal 30,000 hours dataset. There are 12 testsets in this task. Let the averaged CER of the 8 near field tests be \textit{Testset Near}, the averaged CER of the 4 far-field test be \textit{Testset Far}. We compare three types of systems, including \textit{DFSMN-CTC}, \textit{Streaming Base} and \textit{Streaming Large}. For the \textit{DFSMN-CTC} hybrid system, we train DFSMN-CTC systems with 50 DFSMN-layers \cite{zhang2018acoustic}. The language model(LM) used in our hybrid system is a 5-gram language model. For the RNN-T system, we train the system in an end-to-end manner with a vocabulary size of 13,000, which has 7000 Chinese chars and 6000 English BPE subwords \cite{sennrich2015neural}.

As the results shown in Table 2, the \textit{StreamingBase} can get comparable results to our well-trained hybrid baseline system both on \textit{Testset Near} and \textit{Testset Far} with CER 6.80\% and 14.13\%, respectively. Note that our \textit{Streaming Base} has only 46M parameters and does not need any external language models. It has less footprint to deploy this system on devices comparing to the hybrid system. Our \textit{Streaming Large} further improves the performance and gets CER 6.14\% and 12.70\%d on \textit{Testset Near} and  \textit{Testset Far}, respectively. With the same model complexity, our RNN-T system outperforms our hybrid system by relative 9.0\% and 9.2\% improvement on \textit{Testset Near} and \textit{Testset Far}, respectively.

\begin{table}[t]
  \centering
  \small
  \label{table2}
  \begin{tabular}{cccc}
    \toprule
    \multicolumn{1}{c}{\textbf{Model}} & 
    \multicolumn{1}{c}{\textbf{Latency}}  & 
    \multicolumn{1}{c}{\textbf{Param.}}  & 
    \multicolumn{1}{c}{\textbf{Test Near/Far}}\\
    \midrule
    DFSMN-CTC  & 400ms & 110M & 6.75/13.98  \\
    \midrule
   Streaming Base & 400ms & 46M &  6.80/14.13  \\
    \midrule
    Streaming Large & 400ms & 110M & 6.14/12.70  \\
    \midrule
  \end{tabular}
  \caption{Experiment results on 30000-hour-industry task.}
\end{table}

\section{Conclusion}
In this paper, we observed the issue of huge gradient variance during RNN-T training. To address the issue, we propose a normalized jointer network that performs stable convergence during training. We also improved our RNN-T system with a masked conformer encoder network and a transformer-XL predictor network. On the AISHELL-1 dataset, our proposed RNN-T system achieves CER of 6.15\% and 5.37\% for streaming and non-streaming speech recognition without external language models. We further compared our RNN-T system to an internal well-trained hybrid system on a 30000-hour-industry dataset and get 9\% relative improvement.


\bibliographystyle{IEEEbib}
\bibliography{main}

\end{document}